# Different reverse leakage current transport mechanisms of planar Schottky barrier diodes(SBDs) on Sapphire and GaN Substrate

Xiao Wang,[1,2] Zhi-Yu Lin,[4] Yu-Min Zhang,[3] Guo-Qiang Ren[2,a)] and Ke Xu[2,3,4,a)]

AFFILIATIONS:

[1]School of Nano-Tech and Nano-Bionics, University of Science and Technology of China, Hefei, 230026, Anhui, China

[2]Suzhou Institute of Nano-tech and Nano-bionics, Chinese Academy of Sciences, Suzhou, 215123, Jiangsu, China

[3]Suzhou Nanowin Science and Technology Co, Ltd., Suzhou, 215123, Jiangsu, China

[4]Shenyang National Laboratory for Materials Science, Jiangsu Institute of Advanced Semiconductors, Suzhou, 215123, Jiangsu, China

[a] Authors to whom correspondence should be addressed: gqren2008@sinano.ac.cn and kxu2006@sinano.ac.cn

**Abstract:** The effects of different substrates on the off-state leakage current in gallium nitride (GaN) planar diodes are experimentally demonstrated and studied by analyzing temperature-dependent current–voltage characteristics. The two devices exhibit different leakage mechanisms despite being subjected to the same process conditions. The device with a sapphire substrate shows a more intricate leakage mechanism, implying the presence of additional leakage channels compared to the device with free-standing substrates. The electrical leakage characteristics in the planar GaN-on-sapphire device (referred to as utemp-SBD) and GaN substrate device (referred to as usub-SBD) Schottky barrier diodes (SBD) can be described by two distinct processes as the reverse bias gradually increases: device 1. the thermionic emission, variable range hopping (VRH), Frenkel–Poole (FP) emission, and space-charge-limited conduction (SCLC) model; device 2. the thermionic emission (TE), variable range hopping (VRH), and Frenkel–Poole (FP) emission mechanisms. It is noteworthy that in device 1, there is a rapid increase in leakage current within the voltage range of -36 to -40 V, primarily driven by the SCLC mechanism. This effect arises due to the presence of intrinsic traps in GaN grown on a heterogeneous substrate. Additionally, in the voltage range of -7 to -50 V (348 K to 448 K), the leakage mechanism shifts from FP emission to the VRH model, possibly due to variations in the Schottky barrier height.

This study provides an in-depth analysis of the leakage mechanisms observed on different substrates and offers valuable insights for the design of planar gate transistors to minimize or avoid the occurrence of leakage channels.

# Introduction

As the third generation semiconductor material, gallium nitride material has great performance advantages, including high saturated electron drift velocity, high theoretical breakdown field strength, physical stability, large band-gap width and adjustable bandwidth [1-3]. Gallium nitride-based power devices with high output power, high frequency, excellent thermal management and stable dynamic characteristics have attracted much attention [4-10]. These devices can be classified into two primary categories: vertical and planar structures. Vertical power devices possess advantages in terms of on-resistance, breakdown voltage, and device size design [11], however the high cost of free-standing substrate limits its development. The current commercial GaN devices mainly focus on Schottky barrier diode (SBD) and high electron mobility transistor (HEMT) [11]. However, SBDs exhibit a larger leakage current due to their low barrier height compared to other power devices, and the Schottky gate leakage in HEMT devices remains a critical parameter for improving device reliability [12,13]. In the field of microwave applications, gate leakage significantly impacts the noise performance of AlGaN/GaN high electron mobility transistors (HEMTs) [14]. In power devices, leakage current results in additional power losses and reduces device efficiency [15]. Dislocations serve as the primary pathway for leakage current [16,17]. Due to its low dislocation density, the free-standing GaN substrate has the great potential to suppress leakage and enhance the performance of GaN-based devices. In addition, the application of GaN power devices has covered the low, medium and high voltage stages, and they operate in high-temperature conditions. Therefore, it is necessary to systematically measure the temperature-related current and voltage characteristics of Schottky barrier diodes and study their leakage conduction mechanisms.

The conduction mechanism of planar GaN SBD leakage current on different substrates has been limited in previous studies. Zhang et al. reported that variable range hopping (VRH) on gallium nitride vertical diodes is the primary reverse leakage mechanism on all Si, sapphire and gallium nitride substrate [18,19]. Recently, Guo et al. found that the main leakage mechanism varies at different voltages and also differs between devices with and without implantation termination [20]. Chen investigated the leakage current transmission mechanism of quasi-vertical silicon Schottky barrier diodes (SBDs) under low and high reverse bias and different temperatures [21]. A few studies on the leakage conduction mechanism of Schottky diodes primarily focus on GaN-on-Si vertical and quasi-vertical SBDs [21].

In this work, we reveal the different reverse bias leakage conduction mechanisms of the planar GaN-on-sapphire (utemp-SBD) and GaN substrate (usub-SBD) Schottky barrier diodes. We provide a systematic and comprehensive explanation of the physical mechanisms underlying the different leakage conduction mechanisms observed on the two substrates.

## Device Structure and Fabrication

As shown in Fig. 1(a) and (b), the utemp-SBD consists of a 4.5-um-thick high-quality N-drift layer deposited on a sapphire substrate by metal-organic chemical deposition (MOCVD) ($N_D$-$N_A$~$1.7 \times 10^{16}$ cm$^{-3}$, dislocation density is $1 \times 10^9$ cm$^{-2}$), usub-SBD is fabricated by directly depositing electrodes directly on an undoped substrate of GaN ($N_D$~$5 \times 10^{16}$ cm$^{-3}$, Dislocation density is $9.1 \times 10^6$ cm$^{-2}$) grown via Hydride vapor epitaxial deposition (HVPE). Ti/Al/Ni/Au (20/130/50/150 nm) Ohmic electrode were deposited by electron beam evaporation and annealed at 870°C for 30 s in an N2 ambient. The Ni/Au (50/130 nm) were deposited to form Schottky electrode. Both devices do not have any terminal structure or surface treatment. The optical diagram of the device is shown in Fig. 1(c). Usub-SBD I-V curve is shown in Fig. 1(d), the device has good Schottky I-V curve characteristics and the breakdown voltage of the device on the substrate is about 80 V.

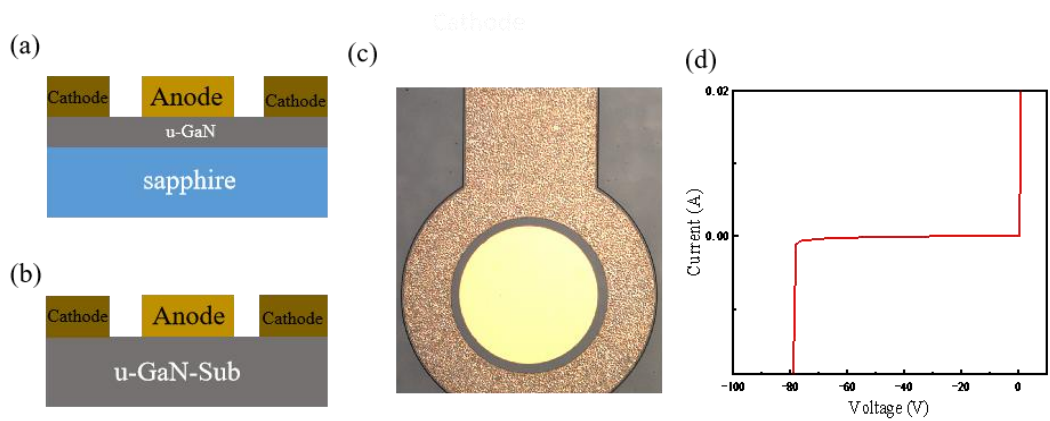

Fig.1(a)(b) Schematic diagram of the planar GaN-on-sapphire (utemp-SBD) and GaN substrate (usub-SBD) Schottky barrier diodes; (c) Optical diagram of the device; (d) Schottky rectification characteristic IV curve.

## Results and Discussion

Fig. 2(a) shows the linear-scale plot of the forward I–V curves of both devices. In general, the forward conduction performance of the Schottky contact is dominated by the thermionic emission (TE) mechanism, which can be expressed as [20]

$$J = J_S \times exp(\frac{qV}{\eta kT})[1 - exp(\frac{-qV}{kT})] \qquad 1)$$

$$J_s = A^*T^2 exp(-\frac{q\phi_B}{kT}) \qquad 2)$$

where J, $J_S$, k, T, q, V, A* and $\phi_B$ are the current density, saturation current density, Boltzmann's constant, temperature, electron charge, voltage across Schottky junction, Richardson's constant and SBH, respectively. The extracted Schottky barrier height (SBH) $\phi_B$ of the utemp-SBD and usub-SBD are 0.62eV and 0.89eV at a forward current density of 1A/cm$^2$, respectively. Under The same process conditions and similar doping concentration, the utemp-SBD shows that lower Schottky barrier height may be related to high dislocation density. Impurity aggregation near the dislocation leads to a higher concentration of donor surface states, the reduced energy level of the localized conduction band is the primary element leading to the observed lower Schottky barrier height [22,23]. Through the variable temperature I-V test, we extracted the barrier heights at different temperatures (298-448 K, step is 25 K), as shown in Fig. 2(b). As the temperature increases from 298 K to 443 K, the barrier height of utemp-SBD and usub-SBD increases from 0.88eV to 0.96eV, and from 0.66eV to 0.86eV, respectively. The ideality factor of the utemp-SBD decreases from 1.6 to 1.4, whereas the ideality factor of the usub-SBD remains nearly unchanged. Thus, compared to the usub-SBD, the utemp-SBD exhibits stronger temperature-dependent characteristics, indicating greater heterogeneity in barrier distribution [24,25]. This can be attributed to the fact that the Schottky contact interface with a higher proportion of dislocations has a larger barrier height difference compared to the normal interface.

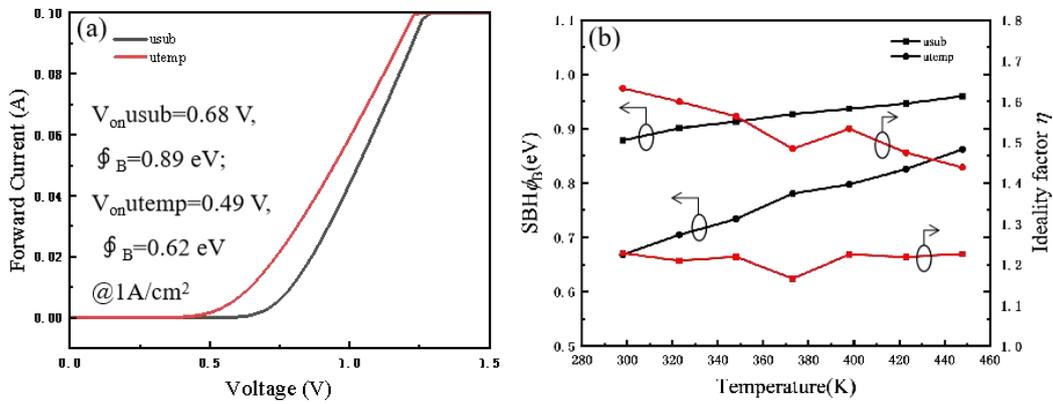

Fig. 2(a) Forward I-V characteristics in a linear scale; (b) Extracted SBH and η versus T using single-temperature I-V method.

In Fig. 3(a), we conducted random testing of the reverse leakage characteristics of two devices of the same size on different substrates. It is observed that the device on the GaN substrate exhibits a more uniform leakage current compared to the device on the sapphire substrate. Fig.3(b)(c) shows the temperature-dependent reverse I-V characteristics of SBDs on both substrates. The curve trend can be divided into several voltage regions. For the utemp-SBD, region I is close to zero bias ($V_R$<1.5), region II ranges from -1.5 V to -12 V, region III spans from -12 V to -36 V, and region IV covers the range from -36 V to -40 V. Usub-SBD region I is close to zero bias (VR<0.5), while region II is from -0.5 to -50 V. utemp-SBD shows a more complex leakage current conduction mechanism with more leakage channels. Leakage currents in quasi-vertical and all-vertical structures are primarily caused by dislocations and defects in conductive layers [26 27]. However, it can be seen from the above results that the dislocation density and defect in the planar structure also have significant effects on the leakage current, because the biggest difference between the two devices in this work is the defect density in GaN.

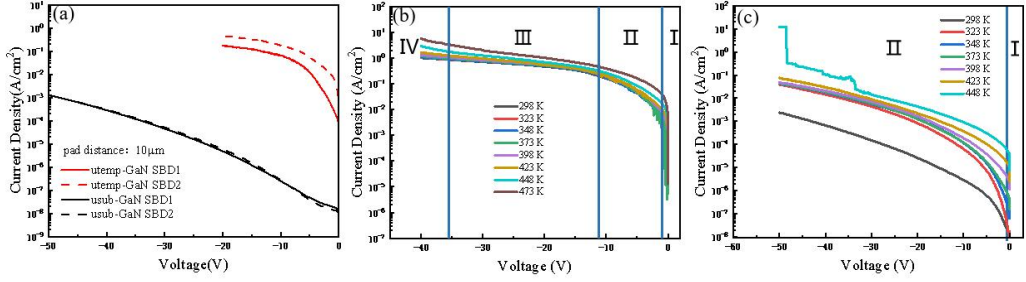

Fig. 3(a) Reverse I-V characteristics on different substrate at room temperature; (b)(c) Temperature-dependent reverse I-V characteristics of the fabricated utemp-SBD and usub-SBD with T varying from 298 to 448 K.

As shown in Fig. 4(a) and (b), it can be observed that when the reverse voltage of the utemp-SBD is below 1.5 V and the usub-SBD is below 0.5 V, there is a linear relationship between LnJ (natural logarithm of current density) and V (voltage) at different temperatures (ranging from 298 K to 443 K). This linear relationship indicates that the dominant conduction mechanism of the leakage current is the thermionic emission (TE) model, which can be described by the equation [21,29]

$$\ln J = \ln(A^*T^2) - \frac{q\phi_B}{kT} + \frac{qV}{\eta kT} \qquad 3)$$

Where η is ideality factor, The well-fitted results verify that the TE conduction is the dominant leakage mechanism [20].

When the reverse voltage of the utemp-SBD is between 1.5 V and 12 V (region II, corresponding to an electric field intensity of up to 0.1 MV/cm), the I-V curves measured at various temperatures exhibit excellent agreement with the theoretical fitting results based on the variable range hopping (VRH) model, described by the equation [21,28]

$$J_{\text{VRH}} = J(0)\exp(C\frac{qEa}{2kT}(\frac{T_0}{T})^{-1/4}) \qquad 4)$$

the linear relationship between lnJ and E [Fig. 4(c)] indicates that the reverse leakage is mainly dominated by the variable range hopping (VRH) process [27], as shown in Fig. 4(d). For the usub-SBD, when the reverse voltage is between 0.5 V - 7 V (electric field intensity is 0.04 - 0.08 MV/cm), the usub-SBD shows a good linear relationship between lnJ and E, as shown in Fig. 4(e). Meanwhile, the lnJ as a function of $1/T^{0.25}$ obtained from the experimental data under -5 V reverse bias exhibits a linear relationship as shown in Fig. 4(e), which confirms that the VRH conduction is the primary leakage mechanism for the usub-SBD between this region. Based on above observations, VRH mechanism begins to act as the dominant leakage mechanism at different voltage values (utemp-SBD and usub-SBD correspond to 1.5 V and 0.5 V, respectively) for two devices, which should be attributed to the different electric field intensity. Due to the higher carrier concentration and higher Schottky barrier height (SBH) of the GaN substrate, it transitions into the VRH mechanism (region II) at a reverse bias voltage of 0.5 V, with a higher electric field intensity of 0.04 MV/cm (the utemp-SBD reaches this value at -1.5 V). In region II, as the reverse bias increases, the electric field becomes higher, causing the energy band to steepen. This steeper VRH band (EVRH) results in a shorter hopping distance from the Schottky Fermi level to the VRH level in GaN [20]. Based on this theory, under the same reverse voltage, the usub-SBD has a steeper VRH band, indicating that electrons have a shorter hopping distance, making the usub-SBD more prone to the VRH conduction mechanism. This is illustrated in the schematic diagram in Fig. 4(f).

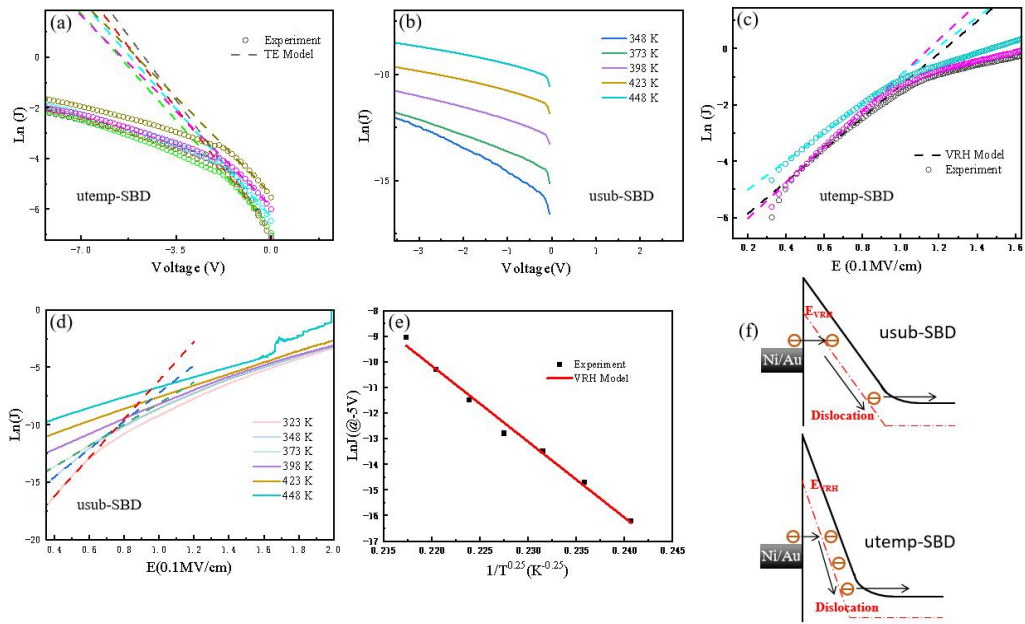

Fig. 4(a)(b) Reverse ln(J) versus V for utemp-SBD and usub-SBD, respectively; ln(J) as a function of applied electric field E; (c)(d) Measured reverse I-V curves and two devices fitting data based on the VRH model; (e) ln(J) versus (1/T)0.25 at reverse voltage of -5; (f)The schematic diagram of usub-SBD and utemp-SBD devices Schottky Fermi level jump to EVRH level at VR=0.5 V

In the low-temperature range (323 K-373 K), when the $V_R$ is raised above 7 V (E ~ 0.08 MV/cm), $\ln(J/E)$ becomes linearly proportional to $E^{1/2}$, as depicted in Fig5 (a), implying that FP emission conduction-related leakage dominates the reverse leakage of the usub-SBD [21]. However, as the temperature increases (398 K-443 K), the dominant leakage mechanism changes from FP emission to VRH mechanism (Fig. 5(a) is observed in combination with Fig. 4(d)). This is different from the previous study [21]. FP emission involves charge carriers being trapped and then emitted to a continuous state, leading to the generation of dynamic current. As the temperature rises, the emission probability of trapped carriers increases [30]. However, in the case of the usub-SBD, the leakage current is mainly dominated by the VRH conduction mechanism, which is an abnormal phenomenon. We suspect that another VRH-related leakage channel, namely surface leakage, annihilates FP emission [31]. One of the possible mechanisms for surface current is two-dimensional variable hopping (2D-VRH) assisted by a high-density surface electronic state in GaN. This transition phenomenon is not observed in the utemp-SBD, as shown in Fig. 5(b). When the reverse voltage is between 12 V and 36 V (region III, E ~ 0.1- 0.17 MV/cm), $\ln(J/E) \propto E^{1/2}$ indicates that the leakage mechanism of the utemp-SBD is dominated by FP emission. Compared to the free-standing GaN substrate, the higher defect density in GaN on sapphire enables FP emission to cover the temperature range.

As the $V_R$ exceeds 36 V (E~0.17 MV/cm) for the utemp-SBD, the leakage current increases more rapidly than the expected value based on the FP emission model as shown in Fig. 5 (c). The well fittings based on the relation of $J / V^n$ imply that the dominant leakage and breakdown mechanisms follow the space-charge-limited conduction SCLC model [20,32]. Under high electric field conditions, the high-density charged intrinsic traps in the space depletion region of the utemp-SBD can modulate the spatial electric field profile and affect the carrier transport process [20,33]. Through simulation in Fig. 5(d), it is observed that different intrinsic defect densities make the two devices exhibit different electric field distributions under the same reverse voltage. In particular, the high-density intrinsic defect state of GaN-on-sapphire leads to a wider depletion region in the utemp-SBD (on the same field intensity scale). This promotes the leakage of a large number of acceptor defect ionization trapping carriers in the space charge region, which is in line with the space-charge-limited conduction (SCLC) leakage mechanism.

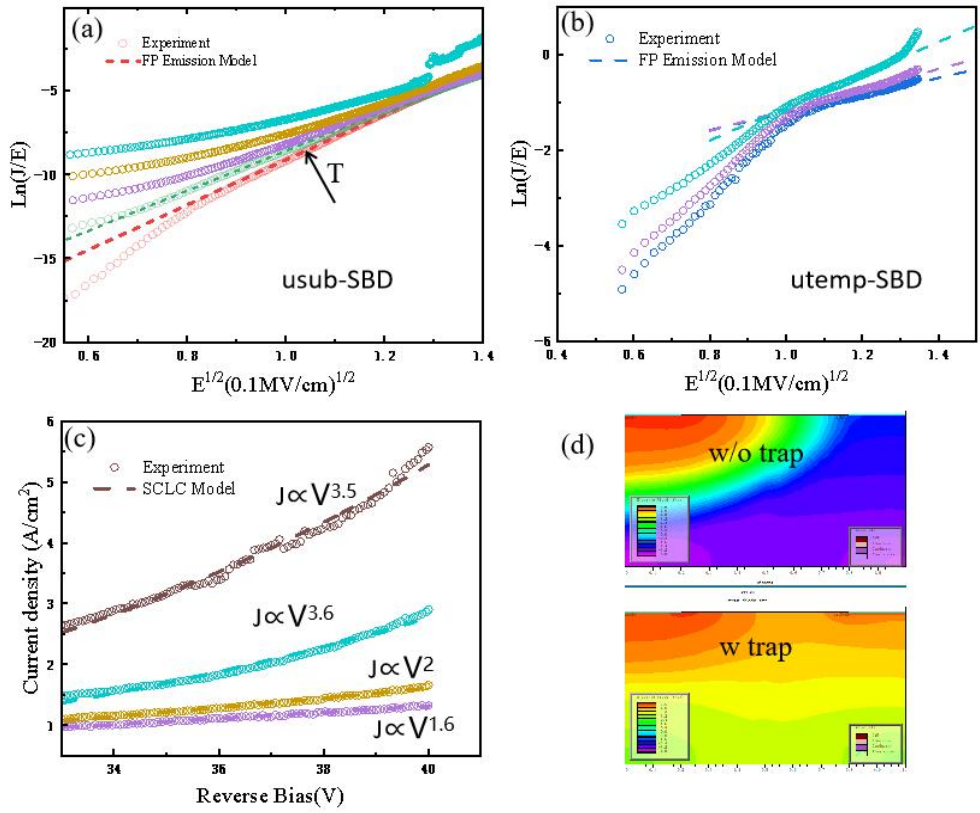

Fig. 5(a)(b) Reverse ln (J/E) versus $E^{1/2}$ for usub-SBD and utemp-SBD, respectively; (c) SCLC model at the high reverse bias, following the relation of $J\text{-}V^n$; (d) Simulated electric field distribution in the usub-SBD and utemp-SBD at the VR of 40 V, respectively

Fig. 6 (a) illustrates the evolution of the various leakage mechanisms as the reverse voltage varies for both the utemp-SBD and the usub-SBD at different temperature. Under near-zero bias voltage, both devices exhibit a TE emission conductance mechanism. As the reverse voltage increases, due to the difference in Schottky barrier height, the larger electric field intensity in the usub-SBD depletion region makes usub-SBD first transform into VRH mechanism. With further voltage increase, the dominant leakage mechanism in the usub-SBD changes to FP emission and the 2D-VRH between the surface states of electrons in the process of temperature rise makes the leakage current dominant mechanism change to VRH. The dominant leakage mechanism in the utemp-SBD changes from FP emission to SCLC mechanism under the influence of intrinsic defects. The primary leakage mechanism model is shown in Fig. 6(b-e).

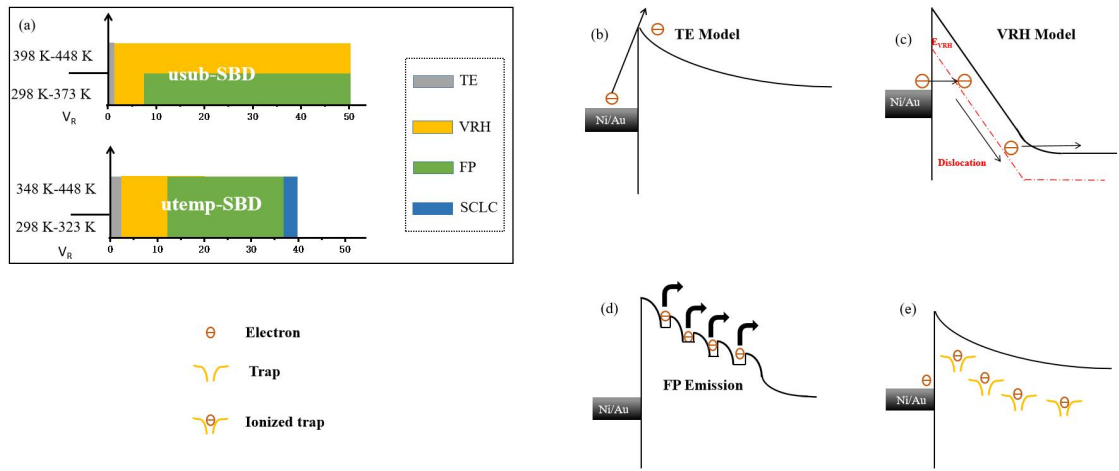

Fig. 6(a) Schematic diagram of various leakage mechanisms for the usub-SBD and utemp-SBD under a reverse bias; The schematic diagram of reverse leakage based on the (b) TE, (c) VRH, (d) FP Emission, (e) SCLC models.

## Conclusion

In summary, this work has conducted a comprehensive analysis of the phenomenon and underlying mechanisms of Schottky barrier diode (SBD) leakage on both free-standing and sapphire substrates through variable temperature current-voltage (IV) testing. The findings indicate that compared to the utemp-SBD, the usub-SBD exhibits a relatively simpler leakage mechanism and lower leakage current. The study analyzes the impact of leakage channels on the leakage mechanism of the device and provides a feasible analysis scheme for suppressing device leakage. These insights are valuable for the design and development of GaN-based power devices, especially in optimizing the performance and reducing leakage in Schottky barrier diodes. By understanding the different leakage mechanisms and their dependence on substrate choice and operating conditions, engineers and researchers can make informed decisions and design strategies to enhance device efficiency and reliability. This work contributes to the overall understanding of leakage mechanisms in GaN-based devices and provides guidance for future device design and optimization efforts.